\newif\ifdraft
 \draftfalse

\newif\ifappendix
\appendixfalse

\newif\iferrorbars
\errorbarstrue

 \documentclass[10pt,conference,twocolumn,letterpaper]{IEEEtran}%

\title{Routing Algorithms for\\ Recursively-Defined Data Centre
  Networks}%
 
\author{\IEEEauthorblockN{Alejandro Erickson and Iain A. Stewart}
\IEEEauthorblockA{School of Engineering and Computing Sciences\\
Durham University, Science Labs, South Road\\
Durham DH1 3LE, U.K.\\
Email: \{alejandro.erickson,i.a.stewart\}@durham.ac.uk}
\and 
\IEEEauthorblockN{Abbas Eslami Kiasari and Javier Navaridas}
\IEEEauthorblockA{School of Computer Science\\
University of Manchester, Oxford Road\\
Manchester M13 9PL, U.K.\\
Email: \{abbas.kiasari,javier.navaridas\}@manchester.ac.uk}} 

\usepackage{amsmath,amsthm,amssymb,graphicx,array,tabularx,url,enumerate,rotating,fancyvrb,longtable, units,mathabx}%
\usepackage{paralist}
\usepackage{algorithmicx,algorithm}
\usepackage{algpseudocode}
\usepackage[nospace,noadjust]{cite}
\usepackage{pgfplots}
\usepackage{pgfplotstable}

\pgfplotsset{
    discard if/.style 2 args={
        x filter/.code={
            \ifnum\thisrow{#1}=#2
                
            \else
            \fi
        }
    }
  }
\pgfplotsset{
    discard if not/.style 2 args={
        x filter/.code={
            \ifnum\thisrow{#1}=#2
            \else
                
            \fi
        }
    }
  }
  
\newtheorem{theorem}{Theorem}[section]%
\newtheorem{lemma}[theorem]{Lemma}%
\newtheorem{definition}[theorem]{Definition}%
\ifdraft
\usepackage{framed}
\fi
\newcommand{\commenta}[1]{\ifdraft\begin{framed}\textcolor{gray}{//Alejandro comment: #1}\end{framed}\fi}
\newcommand{\commentdo}[1]{\ifdraft\begin{framed}\textcolor{red}{//Alejandro todo: #1}\end{framed}\fi}

\newcommand{\bfs}{\texttt{BFS}}

\newcommand{\dimr}[1]{\texttt{DR}\ensuremath{_{#1}}}
\newcommand{\fidcell}{\ensuremath{\mathcal{G}}-\texttt{Cell}}

\newcommand{\gpe}{\texttt{\_E}}

\newcommand{\gpi}{\texttt{\_I}}
\newcommand{\gpt}{\texttt{\_0}}
\newcommand{\scdcn}{SCDCN}
\newcommand{\tar}{\texttt{TAR}}
\newcommand{\tor}{\texttt{TOR}}
\newcommand{\spf}{\texttt{SPF}}
\newcommand{\dfr}{\texttt{DFR}}
\newcommand{\rdn}{RDN}
\newcommand{\ccrdn}{CCRDN}

\newcommand{\gp}{\texttt{GP}}

\newcommand{\pr}{\texttt{PR}}
\newcommand{\nulll}{\textit{null}}

\newcommand\dcellb{$\beta$-\text{DCell}}
\newcommand\dcell{\text{DCell}} \newcommand\ficonn{\text{FiConn}}
\newcommand\pms[2]{\ensuremath{_{#1,#2}}}
\newcommand\dr{\texttt{DCellRouting}} %
\newcommand\set[1]{\{#1\}}%
\newcommand\nflr[2]{\left\lfloor \nicefrac{#1}{#2} \right\rfloor}%

\begin{document}
\maketitle

\commenta{\tableofcontents}

\begin{abstract}
  The server-centric data centre network architecture can accommodate
  a wide variety of network topologies.  Newly proposed topologies in
  this arena often require several rounds of analysis and
  experimentation in order that they might achieve their full
  potential as data centre networks.  We propose a family of novel
  routing algorithms on two well-known data centre networks of this
  type, (Generalized) \dcell\ and \ficonn, using techniques that can
  be applied more generally to the class of networks we call
  completely connected recursively-defined networks.  In doing so, we
  develop a classification of all possible routes from server-node to
  server-node on these networks, called general routes of order $t$,
  and find that for certain topologies of interest, our routing
  algorithms efficiently produce paths that are up to $16\%$ shorter
  than the best previously known algorithms, and are comparable to
  shortest paths.  In addition to finding shorter paths, we show
  evidence that our algorithms also have good load-balancing
  properties.
\end{abstract}

\section{Introduction}
\label{sec:introduction}

The explosive growth of online services powered by data centres (web
search, cloud computing, etc.) has motivated intense research into
data centre network (DCN) design over the past decade and brought
about major breakthroughs.  For example, fat-tree DCNs, introduced in
\cite{Al-FaresLoukissasVahdat2008}, use commodity off-the-shelf (COTS)
servers and switches in a fat-tree (topology), and have resulted in an
evolutionary shift in production data centres towards leaf-spine
topologies, built from COTS hardware.  COTS fat-tree DCNs are not a
panacea, however; for example, fat-trees are difficult to scale.

Research on DCN architecture is ongoing and each new architecture
invites the use of certain classes of topologies. Indirect networks,
where servers are the terminals connected to a switching fabric, are the prevailing 
example. Fat-trees are among the topologies that can be implemented in
indirect network architectures.  A host of alternative topologies can
be implemented as indirect networks, including random regular graphs
(\cite{SinglaHongPopa2012}) and butterfly networks
(\cite{KimDallyAbts2007}).  Likewise, the optical-switch hybrid DCN
Helios (\cite{FarringtonPorterRadhakrishnan2010}) can be seen as an
architecture with the capacity to accommodate a variety of topologies
(both in the wired links as well as in the optical switch itself).
Each architecture sets constraints on the topology in a variety of
ways; for example, by the separation of switching nodes from server
nodes or the number of ports in the available hardware.

The server-centric DCN (\scdcn) architecture, introduced in
\cite{GuoWuTan2008}, accommodates a great variety of network
topologies and has resulted in a number of new DCN designs, both
derived from existing and well-understood topologies in
interconnection networks as well as topologies geared explicitly
towards DCNs (e.g.,
\cite{GuoWuTan2008,LiGuoWu2011,GuoChenLi2013,Abu-LibdehCostaRowstron2010,LiaoYinYin2012,GuoLuLi2009}).

Only dumb crossbar-like switches are used in an \scdcn\ and the
servers are responsible for routing packets through the network.
Therefore, the switches have no knowledge of the network topology and
are only connected to servers.  Servers, on the other hand, may be
connected to both switches and servers.  These parameters, which make
up part of the \scdcn\ architecture, invite sophisticated topologies
from abstractions as graphs, along with accompanying analyses.  We are
concerned primarily with routing algorithms for two well-known \scdcn
s, \dcell\ (\cite{GuoWuTan2008}) and \ficonn\ (\cite{LiGuoWu2011}),
and the topologies called Generalized \dcell\
(\cite{KlieglLeeLi2010,KlieglLeeLi2009}).

We characterise (Generalized) \dcell\ and \ficonn\ as a special case
of \emph{completely connected recursively-defined networks} (\ccrdn),
which we use to develop a classification (which, to our knowledge, is
novel) of all possible routes from server-node to server-node in the
DCNs (Generalized) \dcell\ and \ficonn.  Our main result pertains to a
specific family of routing algorithms, called \pr\ (or
\texttt{ProxyRoute}), which we develop with the primary aim of
improving upon the originally proposed (and best known) routing
algorithms, as regards hop-length.  This goal is achieved with
improvements as high as $16\%$ for certain topologies and paths that
are comparable, in length, to shortest paths. In addition, we give
empirical evidence that the path diversity provided by \pr\ does a
better job of balancing load than \dr.  Hitherto, the only algorithms
for balancing communication load in (Generalized) \dcell\ and \ficonn\
are the adaptive routing algorithms \dfr\ and \tar\ presented in
\cite{GuoWuTan2008,LiGuoWu2011}, so \pr\ is also novel in this
respect.

Two of our instances of \pr\, called \gp\gpi\ and \gp\gpt, exploit the
topological structure of (Generalized) \dcell\ and \ficonn\ in order
to find short paths efficiently by means of an intelligent search (see
Section~\ref{sec:gp-when}) of sub-structures called ``proxies''.  We
then empirically compare the results of our intelligent versions of
\pr\ with a shortest path algorithm, a brute force version of \pr\,
and the routing algorithms that were originally proposed in
\cite{GuoWuTan2008,LiGuoWu2011,KlieglLeeLi2009}.

We give definitions in
Sections~\ref{sec:server-centric}--\ref{sec:rdn}, where we abstract
the DCNs (Generalized) \dcell\ and \ficonn\ as graphs which can be
characterised as \ccrdn s.  Section~\ref{sec:routing} describes
previously known routing algorithms for these DCNs, in the context of
\ccrdn s, and our classification of routes in \ccrdn s is given in
Section~\ref{sec:proxy}, as \emph{general routes of order $t$}.  We
present our main contribution in
Section~\ref{sec:proxy-routing-dcell}: the design of \pr.  Our
empirical work is described and evaluated in
Section~\ref{sec:experiments} and future avenues for research are
identified in the conclusion.

\section{Server-centric DCNs}
\label{sec:server-centric}

Our results and experiments are concentrated on graph theoretical
abstractions of certain \scdcn s.  Therefore, it is appropriate that
we define this abstraction precisely.

An \scdcn\ consists of switches, which act only as crossbars and have
no routing intelligence, and servers.  These components are linked
together, with the only restriction being that a switch cannot be
linked directly to another switch; we assume all links are
bidirectional.  As such, an \scdcn\ is abstracted here by an
undirected graph $G=(W\cup S, E)$, with two types of nodes called
\emph{switch-nodes}, $W$, and \emph{server-nodes}, $S$.  Naturally,
each switch of the \scdcn\ corresponds to a switch-node, $w\in W$, and each
server corresponds to a server-node, $x\in S$.  Each link of the \scdcn\
corresponds to an edge $e$ of $E$, which, for convenience, we shall
also call a link.  The condition that switch-to-switch links are not
allowed implies that $\not\exists (u,v)\in E$ such that $u,v\in W$.
See \cite{Diestel2012} for undefined graph-theoretic terms.

Also relevant to our discussion of routing algorithms in \scdcn s is
the fact that\begin{inparaenum}[(1)]%
\item packets are sent and received only by servers, and
\item packets endure a negligible amount of processing time in each
  switch, compared to the time spent in each server.
\end{inparaenum} The reason for (2) is that we assume the packet is
routed in the server's operating system, either via a table look-up or
computation.  This could be done, e.g., by a dedicated virtual machine
or a specialised hypervisor with the capability to route packets.  In
any case, we may assume that with today's COTS servers, a packet
spends much more time at servers than in switches.

The outcome of (1) is that we need only discuss routing algorithms
that construct paths whose endpoints are server-nodes.  That is, a
\emph{route} on $G$ is a path whose endpoints are server-nodes.  The
outcome of (2) is that a \emph{hop} from server-node to server-node is
indistinguishable from one that also passes through a switch-node.





\section{Recursively-Defined Networks}
\label{sec:rdn}

Our results are concerned with network topologies of a certain form
that have arisen frequently in the area of interconnection networks, and
recently as \scdcn s.

\begin{definition}\label{def:recdef}
  A family $\mathcal X=\{X(h):h=0,1,\ldots\}$ of interconnection
  networks is \emph{recursively-defined\/} if $X(h)$, where $h > 0$,
  is the disjoint union of copies of $X(h-1)$ with the addition of
  extra links joining nodes in the different copies.  We call a member
  of $\mathcal X$ a \emph{recursively-defined network (\rdn)}. A
  family of \rdn s $\mathcal X$ is a \emph{completely-connected \rdn\
    (\ccrdn)} (see, e.g., \cite{ChenHwangSu1998}) if there is at least
  one link joining every copy of $X(h-1)$ within $X(h)$ to every other
  copy.
\end{definition}





\subsection{The DCNs DCell}
\label{sec:dcell}

The DCNs \dcell\ (\cite{GuoWuTan2008}) were the first family of \scdcn
s to be proposed, and their graphs form the family of \ccrdn s
described below.

Fix some $n> 2$.  The graph \dcell\pms 0 n consists of
one switch-node connected to $n$ server-nodes.  For $k\ge 0$, let
$t_k$ be the number of server-nodes in \dcell\pms k n.  For $k>0$,
the graph \dcell\pms k n consists of $t_{k-1}+1$ disjoint copies of
\dcell\pms {k-1} n, labelled $D_{k-1}^i$, for $0\le i \le t_{k-1}$.
Each pair of distinct \dcell\pms {k-1} ns is joined by exactly one
link, called a \emph{level-$k$ link}, whose exact definition is given
below, in terms of the labels of the server-nodes.

Label a server-node of a \dcell\pms n k, for some $k>0$, by
$x=x_kx_{k-1}\cdots x_0$, where $x_{k-1}x_{k-2}\cdots x_0$ is the
label of a server-node in $D_{k-1}^{x_k}$, and $0\le x_0< n$ and
$0\le x_i< g_{k}$ for $i>0$, where $g_{k}=t_{k-1}+1$. The labels of \dcell\pms n k are
mapped bijectively to the set $\set{0,1,\ldots, t_k-1}$ by
$uid_k(x)= x_kt_{k-1}+ x_{k-1}t_{k-2} + \cdots + x_1t_0+x_0$.  Label
and $uid$ are combined in the notation
$[x_k,uid_{k-1}(x_{k-1}x_{k-2}\cdots x_0)]$.

Let $0\le x_k<y_k<t_{k-1}+1$ be the indices of the \dcell\pms {k-1} ns
labelled $D_{k-1}^{x_k}$ and $D_{k-1}^{y_k}$.  A level-$k$ link
connects node $y_k-1$ in $D_{k-1}^{x_k}$ to node $x_k$ in
$D_{k-1}^{y_k}$.  This is the link
$(y_k-1+x_kt_{k-1},x_k+y_kt_{k-1})$.

\subsubsection{Generalized DCell}
\label{sec:generalized-dcell}

The definition of the DCNs \dcell\ generalises readily; see
\cite{KlieglLeeLi2010,KlieglLeeLi2009}.  The key observation is that
the level-$k$ links are a perfect matching of the server nodes in the
disjoint copies of the \dcell\pms {k-1} ns, where every pair of
distinct \dcell\pms{k-1}ns is connected by a link.  Many such
matchings are possible.  A given matching $\rho_k$ which satisfies the
stated properties defines the level-$k$ links and is called a
\emph{$\rho_k$-connection rule} (\cite{KlieglLeeLi2009}).

A \emph{Generalized \dcell\pms k n} inherits the definition of
\dcell\pms k n, for $k\ge 0$, except that the level-$k$ links may
satisfy an arbitrary $\rho_k$-connection rule.  Note that we insist
that there be only one connection rule for each level $k$, so that a
given family of Generalized \dcell s can be specified by a set of
connection rules $\set{\rho_1,\rho_2,\rho_3,\ldots}$.

This is in accordance with  Definition 1 in \cite{KlieglLeeLi2009},
with two exceptions. We model Generalized \dcell\pms 0 n as a
switch-node connected to $n$ server-nodes, rather than modelling it as
$K_n$, and we require $n>2$.

In order to demonstrate the impact of different connection rules on
the routing algorithms presented in Section~\ref{sec:routing}, it
suffices to consider just one connection rule besides the one for
\dcell.  For this purpose, we use \dcellb, defined by the
$\beta$-connection rule given in \cite{KlieglLeeLi2009}.

The $\beta$-connection rule is (perhaps not obviously) as follows: Let
$0\le x_k<y_k<t_{k-1}+1$ be the indices of the \dcellb\pms {k-1} ns
labelled $B_{k-1}^{x_k}$ and $B_{k-1}^{y_k}$.  A level-$k$ link
connects node $y_k-x_k-1$ in $B_{k-1}^{x_k}$ to node $t_{k-1}-y_k+x_k$
in $B_{k-1}^{y_k}$.  This is the link
$(y_k-x_k-1+x_kt_{k-1},t_{k-1}-y_k+x_k+y_kt_{k-1})$.

\subsection{The DCNs \ficonn}
\label{sec:ficonn}

One of the issues with (Generalized) \dcell\pms kn\ is that each
server-node has degree $k+1$.  This requires that each server has
$k+1$ NIC ports, which is not typically the case for COTS servers when
$k>1$.

\ficonn, proposed in \cite{LiGuoWu2011}, is a \ccrdn\ that requires at
most two ports per server; it uses only half of the
\emph{available} server-nodes (those of degree one) in each copy of
\ficonn\pms {k-1} n\ when building \ficonn\pms k n.  This, in turn,
leaves server-nodes of degree one available to build the next level.
We describe \ficonn\ below.

Fix some even $n>3$.  \ficonn\pms 0 n is the network
consisting of one switch-node connected to $n$ server-nodes.  Let $b$
be the number of available server-nodes in \ficonn\pms {k-1} n\ for
$k>0$.  Build \ficonn\pms k n\ from $b/2+1$ copies of \ficonn\pms{k-1}
n, labelled $F_{k-1}^i$, for $0\le i \le b/2$.  From
\cite{LiGuoWu2011} we have that $b/2+1 = t_{k-1}/2^k+1$, so that the
label of a server-node $x$ of a \ficonn\pms k n\ is,
expressed as the $(k+1)$-tuple $x=x_kx_{k-1}\cdots x_0$, where
$x_{k-1}x_{k-2}\cdots x_0$ is a server-node in $F_{k-1}^{x_k}$ and we
have $0\le x_0 < n$, but $0\le x_i < g_k$, where
$g_k=b/2+1=t_{k-1}/2^k+1$ (diverging slightly from the labels in
\dcell).  We have
$uid_k(x)= x_kt_{k-1}+ x_{k-1}t_{k-2} + \cdots + x_1t_0+x_0$ and
$[x_k,uid_{k-1}(x_{k-1}x_{k-2}\cdots x_0)]$ to label server-nodes, once
more.

Let $0\le x_k<y_k<t_{k-1}/2^k+1$ be the indices of the \ficonn\pms
{k-1} ns $F_{k-1}^{x_k}$ and $F_{k-1}^{y_k}$.  A level-$k$
link connects server-node $(y_k-1)2^k+2^{k-1}+1$ in $D_{k-1}^{x_k}$ to server-node
$x_k2^k+2^{k-1}+1$ in $D_{k-1}^{y_k}$.  This is the link
$((y_k-1)2^k+2^{k-1}+1+x_kt_{k-1},x_k2^k+2^{k-1}+1+y_kt_{k-1})$.

\section{Routing}
\label{sec:routing}

\ccrdn s feature a class of routing algorithms that emerges naturally
from their definition, called \emph{dimensional
  routing}. 


\subsection{Dimensional routing}
\label{alg:dimroute}

\begin{definition}
\label{def:routing}
Let $\mathcal X=\set{X(h): h=0,1,\ldots}$ be a family of \ccrdn s, and
let $X_h$ be a copy of $X(h)$, for some fixed $h> 0$.  Let $X_{h-1}^a$
and $X_{h-1}^b$ be disjoint copies of $X(h-1)$ in $X_h$, and let $src$
and $dst$ be nodes of $X_{h-1}^a$ and $X_{h-1}^b$, respectively.
Since $X_h$ is completely connected, there is a level-$h$ link in
$X_h$ incident with a node $dst'$ in $X_{h-1}^a$ and a node $src'$ in
$X_{h-1}^b$.  If $h-1=0$ then either $src=dst'$ or $(src,dst')$ is a
link, and otherwise a path $P_a$ from $src$ to $dst'$ can be
recursively computed in $X_{h-1}^a$.  This same method provides a path
$P_b$ from $src'$ to $dst$ in $X_{h-1}^b$.  A \emph{dimensional
  routing algorithm} on $\mathcal X$ is one which computes paths of
the form $P_a+(dst',src')+P_b$, between any source-destination pair of
nodes in a member of $\mathcal X$, and is denoted
\dimr{\mathcal X}.  A \emph{dimensional route} is one that can be
computed by a dimensional routing algorithm.
\end{definition}

Remarkably (and, perhaps, unfortunately), there are topologies and
source-destination pairs for which no dimensional routing algorithm
computes a shortest path; a notable example is the family of
WK-recursive networks (\cite{VecchiaSanges1988}), for which a shortest
path algorithm is developed in \cite{DuhChen1994}.


\subsubsection{Dimensional routing in (Generalized) \dcell\ and \ficonn}
\label{sec:dcell-routing}

(Generalized) \dcell\ and \ficonn\ are \ccrdn s in which each pair of
disjoint copies of \dcell\pms{k-1} n\ within \dcell\pms k n\ is joined
by exactly one edge.  As such, there is only one choice for the edge
$(dst',src')$, which is computed by the connection rule for level-$h$
links.  Therefore, the connection rules in
Sections~\ref{sec:dcell}--\ref{sec:ficonn} suffice to describe
dimensional routing for these DCNs.

The dimensional routing algorithms for each of these networks serves
as a basis for fault-tolerant and load-balancing routing algorithms
\dfr\ in \cite{GuoWuTan2008}, and \tar\ in \cite{LiGuoWu2011}, and it
is precisely the algorithm called Generalized \dr, given in
\cite{KlieglLeeLi2009}.  The former two are fault and
congestion-tolerant routing algorithms that compute significantly
longer paths, on average, than the dimensional routing algorithms.

\subsection{Proxy Routing}
\label{sec:proxy}

A \emph{general routing algorithm} on a family
$\mathcal X=\set{X(h): h=0,1,\ldots}$ of \ccrdn s is of the following
form. Let $X_h$ be a copy of $X(h)$, for some fixed $h> 0$.  Let
$X_{h-1}^{c_{0}}$ and $X_{h-1}^{c_{t-1}}$ be disjoint copies of
$X(h-1)$ in $X_h$, with $src_{c_{0}}$ and $dst_{c_{t-1}}$ nodes of
$X_{h-1}^{c_0}$ and $X_{h-1}^{c_{t-1}}$, respectively.  Let
$X_{h-1}^{c_{0}},X_{h-1}^{c_{1}}, \ldots, X_{h-1}^{c_{t-1}}$ be a
sequence of copies of $X(h-1)$, where: $c_{0}= a$; $c_{t-1}= b$;
$c_i\neq c_{i+1}$, for $0\le i < t$; and $X_{h-1}^{c_i}$ is disjoint
from $X_{h-1}^{c_j}$ whenever $c_i\neq c_j$.  Let
$(dst_{c_i},src_{c_{i+1}})$ be a link from $X_{h-1}^{c_i}$ to
$X_{h-1}^{c_{i+1}}$, and let $P_i$ be paths in each $X_{h-1}^{c_i}$
from $src_{c_i}$ to $dst_{c_i}$. 

Every routing algorithm computes a path (we shall assume that there
are no repeated nodes) of the form
$P_{0}+(dst_{c_{0}},src_{c_1}) + P_1 + \ldots +
(dst_{c_{t-2}},src_{c_{t-1}}) + P_{t-1}$.

A \emph{general route of order $T$} is one in which $t\le T$ for each
$X(h)$, with $h=0,1,\ldots$ and $t=T$ for at least one of these.  A
\emph{proxy route}, computed by a \emph{proxy routing algorithm}, is a
general route of order $3$ (and a dimensional route is of order $2$).


\subsubsection{\dfr\ for \dcell\  and \tar\ for \ficonn}
\label{sec:dfr-tar}

While we do not provide full details here, we sketch the
proxy-routing-like subroutine that is common to \dfr\
(\cite{GuoWuTan2008}) and \tar\ (\cite{LiGuoWu2011}).  Both \dfr\ and
\tar\ are adaptive routing algorithms which compute paths in a
distributed manner, making decisions on the fly, based on information
that is local to the current location of the packet being routed.

This subroutine computes a part of a proxy route to replace a sub-path
of the intended route.  In particular, a packet may bypass a level $m$
link, $e$, from sub-structure $D_{m-1}^a$ to $D_{m-1}^b$ by re-routing
through a proxy, $D_{m-1}^c$, with $a,b$, and $c$ distinct.  The
decision to bypass is made when the packet arrives at $e$ (or near
$e$, as determined by a parameter in \dfr), and upon its arrival in
$D_{m-1}^b$, the packet is routed directly to its final
destination.

The algorithms \dfr\ and \tar\ produce much longer than \dimr{}, on
average.  The simulations in \cite{GuoWuTan2008} show that \dfr,
although fault-tolerant, computes paths that are over 10\% longer than
the shortest paths, on average, even with as little as 2\% failures.
The maximum length of a route computed by the implementation of \tar\
in \cite{LiGuoWu2011} (Theorem 7) is $2\cdot3^k-1$, whilst it is
$2\cdot 2^{k}-1$ for \dimr{}\ (called \tor\ in \cite{LiGuoWu2011}).
This is reflected in their simulations of random and burst traffic,
where \tar\ computes paths that are 15-30\% longer, on average, than
those computed by \dimr{}.

\commentdo{Say something about how \pr\ improves upon \dfr\ and \tar.}

\section{Proxy routing in \dcell\ and \ficonn}
\label{sec:proxy-routing-dcell}


We propose that proxy routing be used more broadly than it is in \dfr\
and \tar, and with the primary goal of efficiently computing short
paths, rather than fault-tolerance and balancing load,
by applying it in a fundamentally different manner: firstly, we seek
to compute a proxy route at the outset, rather than building the route
piecemeal; secondly, we use this pre-planning in order to find a
proxy route that offers a high degree of savings over the dimensional
route.

One reason for focusing on $t\le 3$ is that visiting each
$X_{m-1}^{c_j}$, for $0<j<t-1$, has an associated cost, and when $m$
is small, as it is when our graphs represent DCNs with a realistically
deployable number of servers, it becomes less likely that general
routes with $t>3$ will be useful.  Furthermore, the methods of
searching for a ``good'' proxy that we explore here may become
impractical for $t>3$, because the search space of potential
(multiple) proxies is much larger.


Henceforth we use \fidcell\ in place of (Generalized) \dcell\ and
\ficonn\ whenever we make statements or arguments that apply to all of
these.

The following lower bound on the hop-length of a general route of
order $t$ is obvious.

\begin{lemma}
  \label{lem:smallb}
  Let $src$ and $dst$ be server-nodes in a \fidcell\pms k n, with
  $k>0$, such that $src$ is in $D_{k-1}^a$ and $dst$ is in
  $D_{k-1}^b$, with $a\neq b$. A general route of order $t$ has length
  at least $2t-3$.  In particular, a dimensional route has length at
  least $1$ and a proxy route has length at least $3$.
\end{lemma}

The remainder of our paper is a comparative empirical analysis of
several versions of \pr, given in Algorithm~\ref{alg:proxy}.

\begin{algorithm}
  \caption{\pr\ for \fidcell\ returns a proxy route if it finds
    one that is shorter than the corresponding dimensional route.}
  \label{alg:proxy}
  \begin{algorithmic}
    \Require $src$ and $dst$ are server-nodes in a \fidcell\pms k n.
    \Function{\pr}{$src,dst,m$} %
    \If{$m>0$ and both $src$ and $dst$ are in the same \break\hspace*{2.5em}copy of
      \fidcell\pms {m-1} n }%
    \State\Return $\pr(src,dst,m-1)$%
    \EndIf%
    \State $D_{m-1}^c\leftarrow \gp(src,dst,m)$.  %
    \If{$D_{m-1}^c=null$} \State\Return \dimr{}$(src,dst)$.%
    \Else%
    \State $D_{m-1}^a\leftarrow$ the \fidcell\pms{m-1}n containing $src$.%
    \State $D_{m-1}^b\leftarrow$ the \fidcell\pms{m-1}n containing $dst$.%
    \State$(a^c,c^a)\leftarrow$ the link from $D_{m-1}^a$ to $D_{m-1}^c$.%
    \State $(c^b,b^c)\leftarrow$ the link from $D_{m-1}^c$ to $D_{m-1}^b$.%
    \State\Return
    \begin{equation}
      \label{eq:proxypath}
      \begin{aligned}
        &\pr(src,a^c,m-1)+(a^c,c^a)+\\
        &\pr(c^a,c^b,m-1)+(c^b,b^c)+\\
        &\pr(b^c,dst,m-1).
    \end{aligned}
  \end{equation}
  \EndIf%
    \EndFunction %
  \end{algorithmic}
\end{algorithm}

\subsection{\gp: \texttt{GetProxy}}
\label{sec:gp-when}

\gp\ is the subroutine of \pr\ that computes the proxy used in
Expression~\eqref{eq:proxypath}, if a proxy is to be used.  That is,
\gp\ returns either a proxy sub-\fidcell, $D_{m-1}^c$, or it returns
\nulll.  Obviously, the performance of \pr\ (and its success in
producing a shorter route than \dimr{}) depends on the proxy returned by
\gp\ and how \gp\ is implemented.

Ideally \gp\ would instantly compute a unique proxy sub-\fidcell\
$D_{m-1}^c$, if it exists, such that the proxy route through
$D_{m-1}^c$ is the shortest one possible.  Such an algorithm is
unknown to us.

Our strategy, however, is widely applicable, as regards different
connection rules and path diversity.  Every version of \gp\ that we
explore is of the following form. Let $(src,dst,m)$ be the inputs to
\gp.  If $m=0$, \gp\ outputs \nulll; otherwise, let $m>0$, so that
$src$ is in $D_{m-1}^a$ and $dst$ is in $D_{m-1}^b$, for some $a$ not
equal to $b$.  \gp\ computes a set of \emph{candidate proxies},
$\set{D_{m-1}^{c_0}, D_{m-1}^{c_1}, \ldots, D_{m-1}^{c_{R-1}}}$ (taken
from the set of all potential proxy \fidcell\pms {m-1}ns), and then
finds a $c_i$ for which the path in Expression~\eqref{eq:proxypath} is
shortest (replacing $c$ by $c_i$), by constructing the paths
explicitly.  If the set of candidate proxies is empty, then \gp\
returns \nulll.

The key observation is that we must minimise the number of candidate
$D_{m-1}^{c_i}$s in order to reduce the search space.  Our goal is to
identify and evaluate general techniques towards this end, and not to
catalogue all of the ways to tune \gp.  Some more complicated
techniques are avoided because there is no room to discuss them in
this paper; for example when routing in a \fidcell\pms k n we only
apply \pr\ at the top level, whereas slightly shorter paths can be
obtained, on average, by using proxy routes in the recursive calls to
\pr\ at Expression~\eqref{eq:proxypath}.  Other techniques are avoided
because they are evidently unprofitable; for example, a much larger
search is encountered if \gp\ computes proxy paths for each proxy
candidate.  We describe  three strategies for generating the 
candidate proxies below.

\subsubsection{\gp\gpe\ as an exhaustive search}
\label{sec:gp-exhaustive}

A proxy \dcell\pms {m-1} n~ $D_{m-1}^c$ can be obtained, na\"ively, if
\gp~is implemented as an exhaustive search; that is, we perform the
steps described in Section~\ref{sec:gp-when} for every $c$ in
$\set{0,1,\ldots, t_{m-1}}\setminus\set{a,b}$.  Measuring the length
of each proxy route has an associated cost, but \gp\gpe\ provides the
optimal proxy route with top-level proxies only against which to test
the two strategies given below.

\subsubsection{\gp\gpi\ as an intelligent search}
\label{sec:intelligent}

\begin{figure}[h]
  \centering
  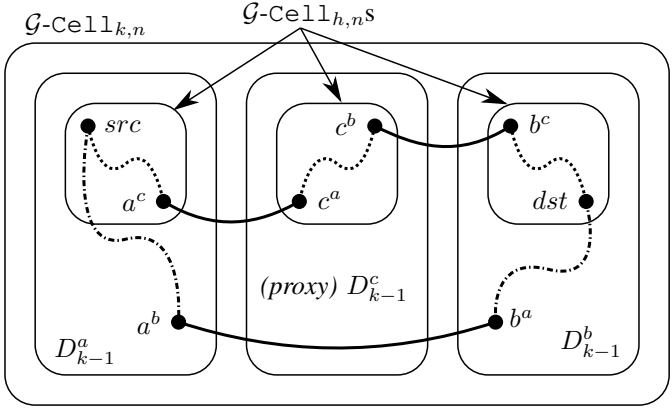
  \caption{Strategy for \gp\gpi, where $h=k-2$, and for \gp\gpt\,
    where $h=0$: select $c$ such that at least one sub-path is
    contained in a \fidcell\pms h n.  Solid arcs represent links, and
    dashed or dotted curves represent paths.}
  \label{fig:rri-figure}
\end{figure}

We propose a general method for reducing the proxy search space, based
on the labels of $src$ and $dst$.  In particular, we look at proxies
$D_{k-1}^c$ whose relationship to $D_{k-1}^a$ and $D_{k-1}^b$ is such
that at least one of the routes computed by the recursive calls to
\pr\ is confined to a \fidcell\pms {k-2} n (see Fig.~\ref{fig:rri-figure}).

\def\svgwidth{\linewidth}

We first give some notation.  Henceforth, let $D_{k}$ be an instance
of \fidcell\pms k n, and let \dimr{}\ be the dimensional routing
algorithm on \fidcell.  For clarity of exposition we describe a method
for selecting a proxy $D_2^c$ when routing in a \fidcell\pms k n, with
$k=3$, but the notation extends to all $k>1$.

Let $src$ and $dst$ be nodes in a \fidcell\pms 3 n, with
$src=a_3a_2a_1a_0$ and $dst=b_3b_2b_1b_0$, so that
$uid_3(src)=t_2a_3+t_1a_2+t_0a_1+a_0$ and
$uid_3(dst)=t_2b_3+t_1b_2+t_0b_1+b_0$. Let $a_3\neq b_3$, and note
that without loss of generality, we may assume $a_3<b_3$.

Our convention for denoting the link between two sub-\fidcell s is as
follows: let $D_2^\alpha$ and $D_2^\beta$ be \fidcell\pms 2 ns and
recall that we may write $[\alpha,uid_2(v)]$ for a node
$v=\alpha v_2v_1v_0$ in $D_2^\alpha$, where
$uid_2(v)=t_1v_2+t_0v_1+v_0$.  Let
$([\alpha,\alpha^\beta],[\beta,\beta^\alpha])$ be the link from
$D_2^\alpha$ to $D_2^\beta$, with
$\alpha^\beta=\alpha^\beta_2\alpha^\beta_1\alpha^\beta_0$, and
similarly for
$\beta^\alpha=\beta^\alpha_2\beta^\alpha_1\beta^\alpha_0$.

\gp\gpi\ builds its set of proxy candidates on the condition that the
source and destination are not near to each other.  Let $a=a_3$ and
let $b=b_3$.  \gp\gpi\ outputs \nulll\ if $[a_3,a^b]$ is a server-node of
$D_1^{a_2}$ or $[b_3,b^a]$ is  a server-node of $D_1^{b_2}$.  That is,
when $a_2= a^b_2$ and $b_2= b^a_2$.  

Provided the above condition is avoided, we then select a proxy
$D_{2}^c$ to be a candidate, when $c$ is such that one of the three
sub-paths, $\pr(src,[a,a^c])$ or $\pr([c,c^a],[c,c^b])$ or
$\pr([b,b^c],dst)$, is short; specifically, if at least one of the
three sub-paths is contained inside a single \fidcell\pms 1 n.  That
is, $c$ satisfies at least one of the following three properties (in a
non-trivial way; see discussion below):
\begin{align}
  \label{eq:d1}
    src \text{ and } [a,a^c] &\text{ are in the same } D_1: a_2=a^c_2\\
  \label{eq:d2}
  [c,c^a] \text{ and } [c,c^b] &\text{ are in the same } D_1:  c^a_2=c^b_2\\
  \label{eq:d3}
  [b,b^c] \text{ and } dst &\text{ are in the same } D_1:
                             b_2=b^c_2,
\end{align}
where $a^c_2=\left\lfloor\nicefrac{a^c}{t_1}\right\rfloor$ and
similarly for $c^a_2$ and $b^c_2$.  Clearly for any \fidcell\ we can
verify whether a proxy candidate $D_2^c$ satisfies one (or more) of
the Properties~\eqref{eq:d1}--\eqref{eq:d3}, since the numerators are
computed directly from the various connection rules of each \fidcell.
However, we wish to compute the set of values $c$ which satisfy
Properties~\eqref{eq:d1}--\eqref{eq:d3} in constant time.

The floor function yields that
$\left\lfloor\nicefrac{a^c}{t_1}\right\rfloor=a_2$ if, and only if,
$a_2t_1\le a^c < (a_2+1)t_1$.  It happens that for our connection
rules (see Sections~\ref{sec:rdn}), $a^c$ is piecewise linear (as a
function of $c$), and similarly for $b^c$, $c^a$, and $c^b$, with
exactly three cases: namely, $c_3<a_3<b_3$; $a_3<c_3<b_3$; and,
$a_3<b_3<c_3$ (where the case $b_3<a_3$ is treated by swapping $src$
and $dst$).  As a result of this, the set of values $c$ which satisfy
Properties~\eqref{eq:d1}--\eqref{eq:d3} can be computed very
efficiently for our connection rules as the union of, at most, a
constant number of intervals (see Table~\ref{tab:alts}). Note that for
the connection rules explored in this paper Property~\eqref{eq:d2} is
redundant because it does not narrow the search space; for certain
pairs $(a,b)$, all $c$ satisfy Property~\eqref{eq:d2},  while no
$c$ satisfies it for other pairs.

\begin{table*}[t]
  \centering
  \begin{tabular}{r@{ to }l|r@{$=$}lr@{$=$}lr@{$=$}l}
    \multicolumn{2}{c|}{route $\backslash$ $c$} & \multicolumn{2}{c}{$c_3<a_3<b_3$}& \multicolumn{2}{c}{$a_3<c_3<b_3$}& \multicolumn{2}{c}{$a_3<b_3<c_3$}\\
    \hline
    $a_3a_2a_1a_0$&$[a,a^c]$&$\nflr{c_3}{t_1}$&$a_2$ & $\nflr{c_3-1}{t_1}$&$a_2$&  $\nflr{c_3-1}{t_1}$&$a_2$\\
    $[c,c^a]$&$[c,c^b]$&$\nflr{a_3-1}{t_1}$&$\nflr{b_3-1}{t_1}$ & $\nflr{a_3}{t_1}$&$\nflr{b_3-1}{t_1}$ &$\nflr{a_3}{t_1}$&$\nflr{b_3}{t_1}$\\
    $[b,b^c]$&$b_3b_2b_1b_0$&$\nflr{c_3}{t_1}$&$b_2$ & $\nflr{c_3}{t_1}$&$b_2$  &$\nflr{c_3-1}{t_1}$&$b_2$ \\    
  \end{tabular}
  \caption{Properties~\eqref{eq:d1}--\eqref{eq:d3} applied to \dcell\pms 3 n.}
  \label{tab:alts}
\end{table*}

For the case $k=3$ and the connection rules for \dcell, \dcellb, and
\ficonn, \gp\gpi\ considers a small set with around $t_1$ or $2t_1$
candidate proxies.  
More generally, a close inspection of Properties~\eqref{eq:d1} and
\eqref{eq:d3} reveals that they each yield exactly $t_2$ (possibly
disjoint) candidate proxies for Generalized \dcell\ and at most $t_1$
candidate proxies for \ficonn.  Due to space constraints we omit a
full discussion of this, but we remark that a better understanding of
this aspect of proxy routes may shed light on the sophisticated
relationship between the connection rule and various distance metrics on
\fidcell.


\subsubsection{\gp\gpt\: level-$0$ proxy search}
\label{sec:tight}

We note that for a \fidcell\pms k n, with $k=2$, the proxy candidates
$D_{1}^c$ computed by \gp\gpi\ are simply those for which $a^c$ is in
the same copy of \fidcell\pms 0 n as $src$ or $b^c$ is in the same
copy of \fidcell\pms 0 n as $dst$ or $c^a$ and $c^b$ are in the same
copy of \fidcell\pms 0 n.  \gp\gpt\ mimics \gp\gpi, but computes the
set of proxies that satisfy at least one of the aforementioned
properties, in place of Properties~\eqref{eq:d1}--\eqref{eq:d3}.  It
is applied only to \fidcell\pms k n with $k>2$.

\commenta{removed section: \gp\gpi\ for the $\rho_k$-connection rule}

\subsubsection{Implementation notes}
\label{sec:implementation-notes}

The savings in hop-length and the benefit to load-balancing come at
the cost of searching proxy candidates, whose number is given by
$\bar p$ in Fig.~\ref{fig:plot-sharp}.  For each proxy candidate $c$,
the lengths of sub-paths $\pr(src,[a,a^c])$ or $\pr([c,c^a],[c,c^b])$
or $\pr([b,b^c],dst)$ must be computed; hence the reason for devising
\gp\gpi\ and \gp\gpt\ with the object of minimising $c$.  Once \gp* is
``tuned'' to suit a certain application and network size, however,
there are several choices for how it can be implemented.  How exactly
this is done depends on the size of the network and the nature of the
application, but we shall remind ourselves of some of the available
tools.

The most na\"ive method is to compute the route at the source-node, by
computing the candidate paths explicitly, and measuring their length,
however, other methods such as table look-ups must to be considered.

\gp\gpi, in particular, leverages the fact that \fidcell\pms k ns grow
double-exponentially in $k$ in order to find proxy candidates
$D_{k-1}^c$ that are linked to the same copy of \fidcell\pms {k-2} n
as $src$ or $dst$. This has a secondary benefit; namely,
\fidcell\pms{k-2}n (and even \fidcell\pms{k-1}n) is small, relative to
\fidcell\pms k n, and this makes table look-ups feasible for storing
the lengths of paths within each copy of \fidcell\pms{k-2}n, and
possibly within each copy of \fidcell\pms{k-1}n.  The whole table must
be replicated at each server-node to be used this way, but this is
still much smaller than storing every $(src,dst)$-pair.  For example,
there are $24,492^2=599,858,064$ such pairs in \dcell\pms 3 3, and
$g_3t_2^2=157*156^2=3,820,752$ pairs confined to sub-\dcell\pms 2 3s,
and $g_3g_2t_1^2=157*13*12^2=293,904$ pairs confined to sub-\dcell\pms
1 3s (see Table~\ref{tab:properties}).

In addition to table look-ups, we also leverage the fact that paths
are computed for flows, rather than packets, and in certain
applications may be re-used for multiple flows among a set of
server-nodes that is small, relative to the entire network.  In
addition, each time we compute a proxy path, we may identify multiple
viable proxies (the context of the application and network size
defines what this means), and hence, path diversity comes at no extra
cost.  We may choose from several paths at random, send a probe packet
to explore the loads and possible faults on each path before sending a
larger flow, or remember proxies for common and recent destinations.

\section{Experiments}
\label{sec:experiments}

\subsection{Experimental setup}
\label{sec:setup}

We compare up to five different routing algorithms for various
\fidcell s.  They are: \dimr{}; shortest paths, computed by a breadth
first search (\bfs); \pr\ with \gp\gpe; \pr\ with \gp\gpi; and, \pr\
with \gp\gpt.  Each routing algorithm (for a given DCN) is tested with
the same $10,000$ input pairs, $(src,dst)$.  The estimated standard
error of the mean is computed by $s_{\bar x}/\sqrt{trials}$, where
$s_{\bar x}$ is the sample standard deviation and $trials=10,000$.
For our purposes of surveying the effects of different instances of
\gp, this value is negligible, and we therefore omit error bars in
Figs.~\ref{fig:plot-bars}--\ref{fig:plot-sharp}.

For each algorithm we plot
$100(\bar x_{\dimr{}}-\bar x)/\bar x_{\dimr{}}$ in
Fig.~\ref{fig:plot-bars}, where $\bar x$ is the mean hop-length in the
sample of computed routes.  In other words, we plot the percent
savings in hop-length over \dimr{}.  Note that \gp\gpt\ is implicitly
plotted for $k=2$ because it is equivalent to \gp\gpi\ in this case.

We also plot, in Fig.~\ref{fig:plot-sharp}, the mean number of proxies
considered by \gp\gpi\ and \gp\gpt\, denoted $\bar p_\gpi$ and
$\bar p_\gpt$, respectively, and the mean number of routes
\pr$(src,dst)$ found to be no longer than \dimr{}$(src,dst)$, denoted
$\bar r_\gpi$ and $\bar r_\gpt$, respectively.  Note that
$\bar p\gpi=\bar p\gpt$ for $k=2$ and, as such, this value is
implicitly plotted for $k=2$ in Fig.~\ref{fig:plot-sharp}.

The two histograms in Fig.~\ref{fig:plot-hist-dcell-beta} show the
proportion of links with a given load (number of flows) in \dcellb\pms
3 3, under $1$ million one-to-one communications, generated uniformly
at random; one histogram is for \dimr{} and the other one is for \pr\
with \gp\gpi.

The networks we tested are given with their basic properties in
Table~\ref{tab:properties}, and the details of each version of \gp$*$
are given in Section~\ref{sec:gp-when}.

\begin{table*}[t]
  \centering
\begin{tabular}{llllllll}
DCN & $N$ & $N/n$ & $|E|$ & $d$ & $g_1$ & $g_2$ & $g_3$ \\
\hline
F\pms{2}{36} & $117648$ & $3268$ & $161766$ & $7$ & $19$ & $172$ \\
F\pms{2}{48} & $361200$ & $7525$ & $496650$ & $7$ & $25$ & $301$ \\
F\pms{3}{10} & $116160$ & $11616$ & $166980$ & $15$ & $6$ & $16$ & $121$ \\
F\pms{3}{16} & $3553776$ & $222111$ & $5108553$ & $15$ & $9$ & $37$ & $667$ \\
F\pms{4}{6} & $857472$ & $142912$ & $1259412$ & $31$ & $4$ & $7$ & $22$ \\
F\pms{4}{8} & $37970240$ & $4746280$ & $55768790$ & $31$ & $5$ & $11$ & $56$ \\
D\pms{2}{18} & $117306$ & $6517$ & $234612$ & $7$ & $19$ & $343$ \\
D\pms{2}{43} & $3581556$ & $83292$ & $7163112$ & $7$ & $44$ & $1893$ \\
D\pms{3}{3} & $24492$ & $8164$ & $61230$ & $15$ & $4$ & $13$ & $157$ \\
D\pms{3}{6} & $3263442$ & $543907$ & $8158605$ & $15$ & $7$ & $43$ & $1807$ \\
\end{tabular}
\caption{Properties of the DCNs in our experiments. We use F  to abbreviate \ficonn, and $D$ to abbreviate ($\beta$-)\dcell.}
  \label{tab:properties}
\end{table*}

\subsection{Evaluation}
\label{sec:results}

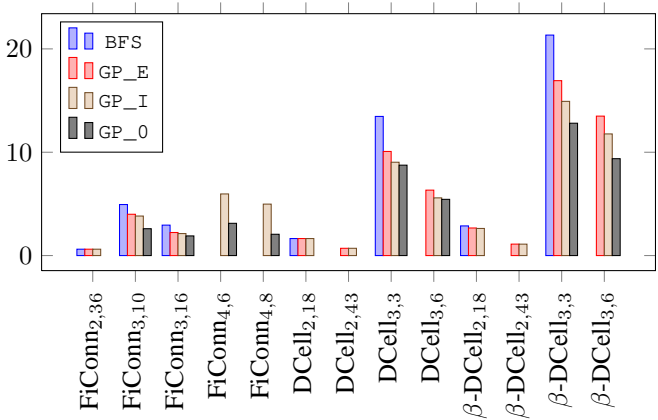
\begin{figure}[ht]
  \begin{tikzpicture}
    \begin{axis}[width=1.1\linewidth,
  height=5cm,
      xtick={0,1,2,3,4,5,6,7,8,9,10,11,12},
      xticklabels={
        \ficonn\pms{2}{36},
      \ficonn\pms{3}{10},
      \ficonn\pms{3}{16},
      \ficonn\pms{4}{6},
      \ficonn\pms{4}{8},
      \dcell\pms 2{18},
      \dcell\pms{2}{43},
      \dcell\pms{3}{3},
      \dcell\pms{3}{6},
      \dcellb\pms{2}{18},
      \dcellb\pms{2}{43},
      \dcellb\pms{3}{3},
      \dcellb\pms{3}{6},
    },
    x tick label style={rotate=90,anchor=east},
    ybar=0pt,
    bar width=3,
    unbounded coords=jump,
    legend entries={
      {\bfs},      
      {\gp\gpe},
      {\gp\gpi},
      {\gp\gpt},
    },
    legend style={font=\footnotesize},
    legend pos=north west,
    ]

    \addplot+[
    ]
    table[
    meta=network,
    x expr=\coordindex,
    y expr=100.0*(\thisrow{mean_dr_list}-\thisrow{mean_bfs})/\thisrow{mean_dr_list},
    ]
    {dat/ispa/allexp_v3_v3.1_bfs.dat};

    \addplot+[
    ]
    table[
    meta=network,
    x expr=\coordindex,
    y=mean_savings_ex_list,
    ]
    {dat/ispa/allexp_v3_v3.1_bfs.dat};

    \addplot+[
    ]
    table[
    meta=network,
    x expr=\coordindex,
    y=mean_savings_int_list,
    ]
    {dat/ispa/allexp_v3_v3.1_bfs.dat};

    \addplot+[
    ]
    table[
    meta=network,
    x expr=\coordindex,
    y=mean_savings_ti_list,
    discard if={k}{2},
    ]
    {dat/ispa/allexp_v3_v3.1_bfs.dat};

\end{axis}     
  \end{tikzpicture}
  \caption{Percent mean hop-length savings over \dimr{}.}
\label{fig:plot-bars}
\end{figure}

\begin{figure}[ht]
  \begin{tikzpicture}
    \begin{axis}[width=.9\linewidth,
  scale only axis,
  unbounded coords=jump,
      xtick={0,1,2,3,4,5,6,7,8,9,10,11,12},
      xticklabels={
        \ficonn\pms{2}{36},
      \ficonn\pms{3}{10},
      \ficonn\pms{3}{16},
      \ficonn\pms{4}{6},
      \ficonn\pms{4}{8},
      \dcell\pms 2{18},
      \dcell\pms{2}{43},
      \dcell\pms{3}{3},
      \dcell\pms{3}{6},
      \dcellb\pms{2}{18},
      \dcellb\pms{2}{43},
      \dcellb\pms{3}{3},
      \dcellb\pms{3}{6},
    },
    x tick label style={rotate=90,anchor=east},
    unbounded coords=jump,
  legend entries={
    {$\bar r_\gpe$},
    {$\bar r_\gpi$},
    {$\bar r_\gpt$},
    {$\bar p_\gpi$},
    {$\bar p_\gpt$},
  },
  legend style={font=\footnotesize},
  legend pos=north west,
  legend columns=1,
  mark size=1.5,
 ]


 \addplot[
 sharp plot,
 mark=square,
 ]
 table[
 meta=network,
 x expr=\coordindex,
 y=mean_n_good_routes_ex_list,
 discard if={k}{4},
 ]
  {dat/ispa/allexp_v3_v3.1_bfs.dat};

 \addplot[
 sharp plot,
 mark=o,
 ]
 table[
 meta=network,
 x expr=\coordindex,
 y=mean_n_good_routes_int_list,
 ]
  {dat/ispa/allexp_v3_v3.1_bfs.dat};

 \addplot[
 sharp plot,
 mark=diamond,
 ]
 table[
 meta=network,
 x expr=\coordindex,
 y=mean_n_good_routes_ti_list,
 ]
  {dat/ispa/allexp_v3_v3.1_bfs.dat};


  
 \addplot[
 sharp plot,
 mark=*,
 ]
 table[
 meta=network,
 x expr=\coordindex,
 y=mean_nproxies_int_list,
 ]
  {dat/ispa/allexp_v3_v3.1_bfs.dat};

 \addplot[
 sharp plot,
 mark=diamond*,
 ]
 table[
 meta=network,
 x expr=\coordindex,
 y=mean_nproxies_ti_list,
 discard if={k}{2},
 ]
  {dat/ispa/allexp_v3_v3.1_bfs.dat};

  



\end{axis}

  
  \end{tikzpicture}
  \caption{Mean number of candidate proxies $\bar p$, and mean number
    of routes no longer than \dimr{}$(src,dst)$, $\bar r$.}
\label{fig:plot-sharp}
\end{figure}
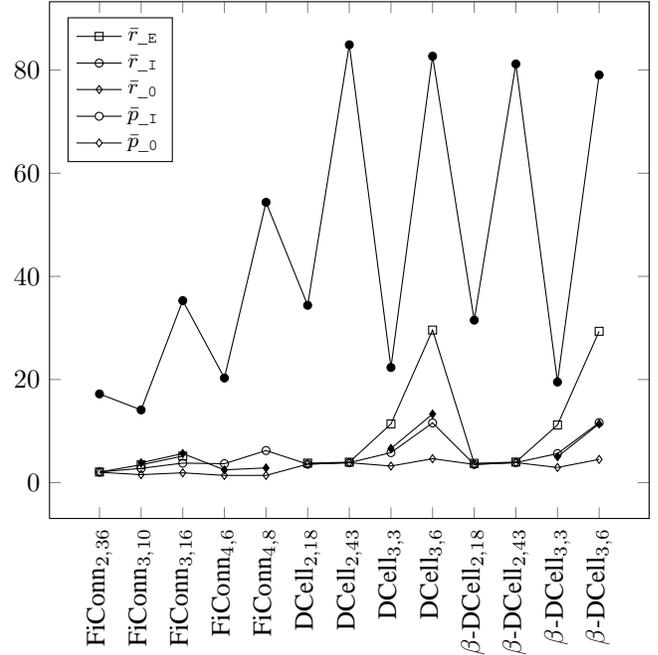

The plots in Fig.~\ref{fig:plot-bars} show that for many \fidcell\
topologies, significant savings in hop-length can be made over
dimensional routes by using proxy routes, depending on the connection
rule, network size, and the parameters $k$ and $n$.  It is immediate
that \gp\gpi\ and \gp\gpt\ retain some good proxies, in relation to
\gp\gpe, which tries all of them.  Furthermore, \gp\gpe\ is comparable
to \bfs.  Fig.~\ref{fig:plot-sharp} tells us how much searching each
of the methods \gp\gpi\ and \gp\gpt\ must do, and how much path
diversity they create, on average.

Note that the means plotted in
Figs.~\ref{fig:plot-bars}--\ref{fig:plot-sharp} hide the success rate
of \pr\ in finding a good proxy path; as a typical example,
\pr$(src,dst)$ is shorter than \dimr{}$(src,dst)$ for approximately
$30\%$ of input pairs when using \gp\gpi\ in \dcell\pms 3 6.

We highlight (and explain, where possible) some of the trends observable
in the plot of Fig.~\ref{fig:plot-bars}:
In general, proxy routes are more effective in \dcellb\pms k * than in \dcell\pms k* and \ficonn\pms k * of comparable size, with fixed $k$, however, even \ficonn\pms k * still sees up to a $6$--$7\%$ improvement.


The apparent weakness of \pr\ in \ficonn\ is partly
explained by the fact that for given $k$ and $n$, there are fewer
proxy \ficonn\pms {m-1}ns to consider at level $m$.  On the other hand
we find that \gp\gpt\ considers fewer than $g_1=6$ proxies for
\ficonn\pms 3{10}, while it considers more than $g_1=7$ proxies for
\dcell\pms 3 6 and \dcellb\pms 3 6.  In addition, there are an equal
number of potential proxy candidates in \dcellb\pms k n and \dcell\pms
k n in general, yet \gp\gpe, \gp\gpi, and \gp\gpt\ invariably consider
more proxy candidates for \dcell\pms k n, only to produce proxy paths
that perform better in \dcellb\pms k n.  We must conclude that the
connection rule and topology (\ficonn\ vs Generalised \dcell)
profoundly impacts the performance of our proxy routing algorithms.
This is somewhat unsurprising, however, since the connection rule and
topology also affect the shortest paths; for example, the mean
distance in \dcellb\pms 3 3 is far shorter than in \dcell\pms 3 3 (see
also \cite{KlieglLeeLi2009}).

Proxy paths in larger networks (when increasing $n$) are worse than those
in smaller networks, for each DCN with fixed $k$; for example
\dcell\pms 3 3 and \dcell \pms 3 6, and also \ficonn\pms 3 {10} and
\ficonn\pms3 {16}.  

A related trend appears to be that for each family of DCNs,
proxy-path-savings increase with $k$, in every version of \gp*; for
example, \ficonn\pms 3{10} and \ficonn\pms 4 6.  The main reason for
this is that the performance of \bfs, relative to \dimr{}, also
increases with $k$, thus providing a greater margin for improvement by
using \pr.


The difference between \gp\gpi\ and \gp\gpt\ grows with $k$ (note that
for $k=2$, they are the same, and hence \gp\gpt\ is not plotted for
$k=2$).  This is because \gp\gpi\ looks for sub-paths within a copy of
\fidcell\pms {k-2}{n}, whereas \gp\gpt\ looks for sub-paths within a
copy of \fidcell\pms{0}{n}, and as the gap between $0$ and $k-2$
increases, \gp\gpi\ considers a larger set of proxy candidates.
Similarly, we explain how the difference between \gp\gpe\ and \gp\gpi\
grows with $k$, but here it is the double exponential growth of
\fidcell\ that contributes extra proxy candidates to \gp\gpe, since
the search space for \gp\gpi\ is proportional to $g_{k-1}$, whereas,
\gp\gpe\ considers exactly $g_k$ proxy candidates (see
Table~\ref{tab:properties}).  Most notably, however, is the fact that
for \fidcell\pms 2*, the performance of \gp\gpe\ is almost identical
to the performance of \gp\gpi; whereas \dcell\pms 2{43} has $g_1=44$,
and $g_2=1893$, our results show that optimal proxies are nevertheless
considered by \gp\gpi\ (and hence, \gp\gpt).

Although \gp* is effective in computing shorter paths and comes fairly
close to \bfs\ (typically over 80\% of the savings are obtained with
\pr), we can confirm that the shortest paths for these topologies are
not, in general, a proxy route of the form we are considering in this
paper as sometimes (e.g.  ($\beta$-)\dcell\pms 3 3) this difference is
considerable.  This was expected, and provides motivation to explore
novel general routing algorithms of order $3$ and higher in future
work.

Another benefit of proxy routing is that it also yields some path
diversity which can be exploited for load balancing and
fault-tolerance purposes. This can be seen in
Fig. \ref{fig:plot-sharp}, where $\bar r$ is the number of distinct
(but not necessarily disjoint) paths considered by \pr$(src,dst)$ that
are no longer than \dimr{}$(src,dst)$.  Additional data must be
studied, however, to determine exactly how $\bar r$ affects the
load-balancing properties of the network.

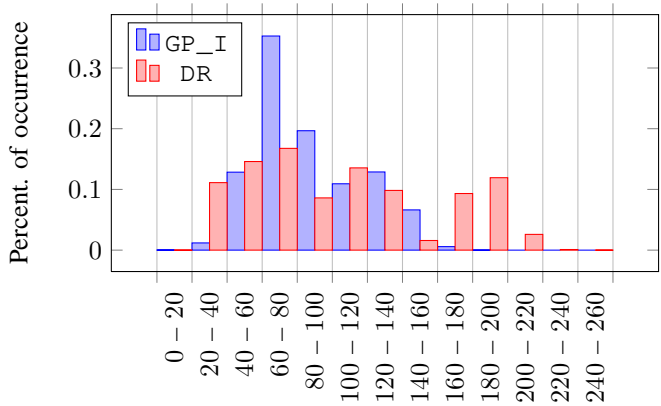
\begin{figure}[ht]

\begin{tikzpicture}
  \begin{axis}[width=\linewidth,
    height=5cm,
    ybar interval,
    xticklabel interval boundaries,
    xtick=data,
    x tick label style={rotate=90,anchor=east},
   ylabel={Percent. of occurrence},
   legend entries={
     {\gp\gpi},
     {\dimr{}},
   },
   legend pos=north west,
   ]

   \addplot table
   {dat/ispa/histograms/DCell_beta_k3_clever_hist_k3n3t1000000s1norm.dat};%
   \addplot table
   {dat/ispa/histograms/DCell_beta_k3_clever_hist_k3n3t1000000splainnorm.dat};%
   
\end{axis}
\end{tikzpicture}

\caption{Normalised histograms showing the proportion of links with a
  given load (number of flows), comparing \dimr{} with \pr\ using
  \gp\gpi\ \dcellb\pms 3 3.}
\label{fig:plot-hist-dcell-beta}
\end{figure}

We computed histograms that show the proportion of links with a given
load, under $1$ million one-to-one communications, plotted in
Fig.~\ref{fig:plot-hist-dcell-beta}.  The histogram for \gp\gpi\ is
shifted left relative to the histogram for \dimr{}, meaning that many
links carry less load than in the same scenario for \dimr{}.  In
addition, the maximum load is reduced (in our sample), suggesting many
\fidcell s have a higher aggregate bottleneck throughput (ABT,
introduced in \cite{GuoLuLi2009}, and closely related to the most
heavily loaded link in the network) with \pr\ than with \dimr{}.

Note that our primary focus is to reduce hop-length and implementation
overheads of \gp, and that we could increase path diversity even more if we were
willing to route on longer paths than \dimr{}$(src,dst)$; we do not do
this here, but will explore this possibility in future research.

\subsection{Significance}
\label{sec:significance}

Various aspects of routing in a DCN depend heavily on the availability
of short one-to-one paths.  For example, minimising latency and energy
usage, and building fault-tolerant and load balancing routing
algorithms.

While there are inherent trade-offs in computing short proxy routes,
there are also multiple benefits: using shorter one-to-one paths in a
DCN reduces the average latency of communications, the aggregate load,
and thereby the energy usage; and, we obtain a non-deterministic path
diversity at no extra cost while computing these paths, which can be
used both adaptively or randomly to deal with faults and congestion,
in addition to forming the building blocks of other fault-tolerant and
load balancing routing algorithms (such as the way \dimr{} is used in
\dfr\ and \tar).  As such, proxy routes are not only a good candidate
for replacing \dimr{} in (Generalized) \dcell\ and \ficonn, they are
also effective at performing some of the functions of the known
adaptive routing algorithms for these networks, namely \dfr\ and \tar,
while simultaneously producing short paths.

\commentdo{expand significance.}

\section{Conclusions and Future Research}
\label{sec:conclusion}

\commentdo{mention that \pr\ does well wrt to \bfs\ and that it
  partially improves on the adaptive algorithms \dfr\ and \tar.}

In this paper we have shown that the topologies of the DCNs
Generalized \dcell\ and \ficonn\ are completely connected
recursively-defined networks. As such, we characterised all possible
routes (with no repeated nodes) on these networks and then proposed
the family of routing algorithms \pr\ to compute proxy routes; that
is, general routes of order 3.  We detailed three instances of this
family, \gp\gpe, \gp\gpi, and \gp\gpt, where each one considers a
number of candidate proxy sub-structures, and selects the optimal
proxy to route through.  We performed an analytical and empirical
comparison between these, shortest paths, and the previously known
dimensional routes, as regards mean hop-length; The main results of
our experiments are that significant savings in hop-length can be made
over dimensional routes by using proxy routes, even with only a
relatively small set of candidate proxies, and that the amount of
savings depends on connection rule, network size, and the parameters
$k$ and $n$.


In future research we will perform a deeper analysis of the DCNs in
question, with two major goals.  The first one, motivated by the fact
that \gp\gpi\ sometimes discards the optimal proxy candidate, calls
for a closer inspection of the topologies.  We want to both find the
optimal proxy candidates, and reduce the size of the search space.

Furthermore, whereas this paper is focused on dimensional and proxy
routing, there may be cases where no shortest path between two
server-nodes is a dimensional route or a proxy route.  Note that
whilst a given shortest path may be found not to be a dimensional or
proxy route, this does not preclude other paths with the same terminal
nodes from being dimensional or proxy routes.  A deeper mathematical
analysis of the DCNs in question may shed light on (1) whether or not
higher-order routing algorithms are needed, and (2) how to compute
optimal routes of this type efficiently.

 \textbf{Acknowledgement }
 This work has been funded by the Engineering and Physical Sciences
 Research Council (EPSRC) through grants EP/K015680/1 and EP/K015699/1.

\bibliographystyle{IEEEtran}
\bibliography{IEEEabrv,bibliography}

\end{document}